\documentclass[aps,twocolumn]{revtex4}

\usepackage{graphicx}% Include figure files
\usepackage{dcolumn}% Align table columns on decimal point
\usepackage{bm}% bold math
\usepackage{amsmath}
\usepackage{graphicx}
\usepackage{listings}
\usepackage[T1]{fontenc}
\usepackage{xcolor}
\usepackage{lmodern}
\usepackage{listings}

\begin{document}

\title{{\bf Evolution of FRW universe in variable modified Chaplygin gas model}}% Force line breaks with \\
%\thanks{As part of Phd under Dr. Sarbari Guha.}%\\

\author{\bf Samarjit Chakraborty$^{1}$}
\author{Sarbari Guha$^{1}$}
\author{D. Panigrahi$^{2}$}
\affiliation{\bf $^1$ Department of Physics, St.Xavier's College, Kolkata 700016, India \\ $^2$ Department of Physics, Sree Chaitanya College, Habra 743268, India}

\maketitle
\section*{Abstract}
 %describe the whole work in short here %

In this paper we study the evolution of the FRW universe filled with variable modified Chaplygin gas (VMCG). We begin with a thermodynamical treatment of VMCG described by the equation of state $P = A\rho-B\rho^{-\alpha}$, and obtain its temperature as a function of redshift $z$. We show that the results are consistent with similar works on other types of Chaplygin gas models. In addition to deriving the exact expression of temperature of the fluid in terms of the boundary conditions and redshift, we also used observational data to determine the redshift at the epoch of transition from the decelerated to the accelerated phase of expansion of the universe. The values of other relevant parameters like the Hubble parameter, the equation-of-state parameter and the speed of sound are obtained in terms of the redshift parameter, and these values are compared with the results obtained from previous works on MCG and other Chaplygin gas models for the various values of $n$ permitted by thermodynamic stability. We assume the present value of temperature of the microwave background radiation to be given by $ T_0 = 2.7 K $, and the parameter $ A $ in the equation of state is taken as $ 1/3 $ since it corresponds to the radiation-dominated phase of the universe. The value of the parameter $\Omega_x$ has been assumed to be $0.7$ in our calculation. Since it is known that the redshift of photon decoupling is $ z\simeq 1100 $, we used this value to calculate the temperature of decoupling.

% in subsequent calcualtions we can use finer values of omega . Redshift of transition will be calculated from these calculations . Temperature of decoupling can be calculated and compared with MCG model using z=1100 . Follow the paper of soares.

% some extra material on this should be added , should not be a copy of soares paper .

KEYWORDS: cosmology; Chaplygin gas; thermodynamic analysis

\section{Introduction}

Einstein's General Theory of Relativity (GTR) revolutionized our understanding of gravity and the structure of space-time. It predicted many new things like the expansion of the universe, space-time singularity, and most recently, the discovery of the gravitational waves was another feather in the cap of GTR. But following the distance measurements of Type Ia supernova \cite{SN1a1,SN1a2,SN1a3,SN1a4,SN1a5}, astronomers came to the understanding that the universe's expansion is accelerated at the present time, an observation which could only be accounted for by the dynamics of a hitherto unknown form of energy, called ``Dark energy'' (DE). Coupled with this was the problem of explaining the observed rotation curve of the galaxies, which lead to the hypothesis of non-baryonic Cold Dark Matter (DM), constituted of particles which are yet to be detected directly. Since then, scientists have proposed a variety of theories to explain these observations. All these models can be broadly classified into two major groups: either one has to change the geometry part of the Einstein field equations to explain these observations, or change the matter-energy part. In an effort to modify the matter part, several researchers proposed the existence of various exotic fluids, a prominent one being the `quintessence', to explain the Dark energy. Dark Matter is gravitationally attractive, being responsible for the clustering of matter in the universe, whereas dark energy is repulsive and responsible for the accelerated expansion of the universe. At the same time, scientists were also looking for a model which could simultaneously explain the mechanism of both the DE and DM. These searches led to the development of the so-called \emph{Chaplygin Gas Cosmology}.

The \emph{Chaplygin gas} (described by the equation of state $P=-B/\rho$) \cite{CG1,CG2}, is an exotic perfect fluid. It explains both the aspects of DE and DM in a simple way and at the same time conforms to the observational data quite well. Several models of Chaplygin gas have been proposed in succession to explain the observational data more accurately. The simplest one is the generalized Chaplygin gas (GCG) \cite{GCG1,GCG2,GCG3,GCG4} with an equation of state
\begin{equation}\label{01}
P=-B/\rho^\alpha,
\end{equation}
where $ B $ is a positive constant and the parameter $\alpha$ takes on values such that $ 0<\alpha\leq 1 $.

The variable Chaplygin gas (VCG) was first proposed by Zhang and Guo \cite{VCG1,VCG2} with the equation of state
\begin{equation}\label{02}
P=-B(a)/ \rho,
\end{equation}
where the constant coefficient $B$ is replaced by a variable coefficient  $ B(a)=B_{0}a^{-n}$. Although it explains two important phases of the evolution of the universe: the dust phase and the present accelerated expansion phase, but it did not capture the earlier radiation-dominated phase of the universe. Hence came the next model: the modified Chaplygin gas (MCG) \cite{MCG1,MCG2} with the equation of state
\begin{equation}\label{03}
P= A\rho-B\rho^{-\alpha},
\end{equation}
where $A$ and $B$ are positive constants. This MCG model has the amazing capability to describe all three evolutionary phases of the universe, starting with the radiation phase (with $A=1/3$), then going through a pressureless phase (dust phase), and then transiting into the present negative pressure phase dominated by dark energy.

Subsequently, in order to explain the observational data even more accurately, researchers came up with more refined models in which the parameter $B$ was assumed to be a function of the scale factor $a(t)$ of the FRW universe. This led to two models, namely, the variable generalized Chaplygin gas (VGCG) and the variable modified Chaplygin gas (VMCG) \cite{VMCG1}. The VMCG equation of state is
\begin{equation}\label{04}
P= A\rho-B(a)\rho^{-\alpha},
\end{equation}
where $B(a)=B_{0}V^{-n/3}$, or $ B(a)=B_{0} a^{-n} $ (for FRW universe). Here the parameters $A$, $B_{0}$ are positive constants and $n$ is also a constant. This model can describe dark energy more accurately because of the extra free parameter $n$ appearing in the equation of state.

Once a cosmological model is proposed, it becomes necessary to examine the viability of such models from the point of view of the corresponding cosmological dynamics, as well as its thermodynamic stability. Several authors have already worked on these aspects (see for example \cite{VMCG2,VGCG3}). Here in this paper, we will deduce the temperature evolution of the FRW universe filled with VMCG as a function of red shift $ z $. We will also use observational data to determine the redshift at the epoch when the transition from deceleration to acceleration happened. We deduced the values of other relevant parameters like the Hubble parameter, the equation-of-state parameter and the speed of sound in terms of the redshift parameter and examined how these values differ from the results obtained from previous works on MCG and other Chaplygin gas models for the various values of $n$ permitted by thermodynamic stability. The temperature of decoupling is calculated with the value of decoupling redshift as $z\simeq 1100$.

\section{Thermodynamic analysis}
The metric corresponding to the flat FRW universe is given by

\begin{equation}\label{05}
ds^{2}=-dt^{2} + a^{2}(t)(dr^{2}+r^{2}d\theta^{2}+r^{2}sin^{2}\theta d\phi^{2}),
\end{equation}
where $ a(t) $ is the scale factor. For the sake of calculations, we have assumed $ V=a^{3} $ for the FRW universe.
The equation of state of the VMCG is
\begin{equation}\label{06}
P= A\rho-B\rho^{-\alpha},
\end{equation}
where $B=B_{0}V^{-n/3}$. Now we have the well known thermodynamic identity $\left(\frac{\partial U}{\partial V}\right)_s = -P $, in which we substitute (\ref{06}) to get
\begin{equation}\label{07}
\left(\frac{\partial U}{\partial V}\right)_s = -A(U/V)+ B_0 V^{-n/3} (V/U)^{\alpha}.
\end{equation}
From this equation (\ref{07}), the energy density is determined accurate up to the order of an integration constant in the form
\begin{equation}\label{08}
 \rho = \frac{1}{a^\frac{n}{1+\alpha}}\left[(1+\alpha)B_0/N + C / a^{3N}\right]^\frac{1}{1+\alpha},
\end{equation}
where $N=(A+1)(1+\alpha)-n/3$, and $C$ is the integration constant which can be an universal constant or a function of entropy $S$. Using the boundary condition i.e. the present day energy density $ \rho_0=\rho(a_0) $ in the above relation (\ref{08}), we can determine the integration constant in terms of $ \rho_0$ and $a_0 $. The resulting expression of energy density for the VMCG in FRW universe as a function of scale factor is
\begin{equation}\label{09}
\rho(a)=\dfrac{\rho_{0}}{a^{\frac{n}{1+\alpha}}}\left[\Omega_x + (a_0^{n}-\Omega_{x})(a_0/a)^{3N}\right]^\frac{1}{1+\alpha},
\end{equation}
where we have defined the dimensionless parameter
\begin{equation}\label{10}
\Omega_x =\dfrac{(1+\alpha)B_0}{N \rho_{0}^{1+\alpha}}.
\end{equation}
Introducing the parameter $ R=(1+A)(1+\alpha) $, and substituting $ n=0 $ in the above equation (\ref{09}), we get the energy density for MCG as
\begin{equation}\label{11}
\rho(a)=\rho_{0} \left[\Omega_x + (1-\Omega_{x})(a_0/a)^{3R}\right]^\frac{1}{1+\alpha}.
\end{equation}
The expression (\ref{09}) for the energy density can also be derived using the field equations for FRW cosmology. Here we have used purely thermodynamic approach and got the same expression. This in fact shows the close relation between GTR and thermodynamics.
% here have to show how i get this relation

We know that the first law of thermodynamics can be written in the form
\begin{equation}\label{12}
TdS=d(\rho/{m})+P d(1/{m}),
\end{equation}
where $ S $ is the entropy per particle, $ \rho $ is the total energy density, $ P $ is the pressure, $ T $ is the temperature in Kelvin, and $ m $ is the particle density in the system. Equation (\ref{12}) can be rewritten as
\begin{equation}\label{13}
dS=(1/Tm)d\rho-(P+\rho)/T m^2 dm.
\end{equation}
This leads us to two thermodynamic relations:
\begin{equation}\label{14}
\left(\dfrac{\partial S}{\partial \rho}\right)_m = 1/Tm,
\end{equation}
and
\begin{equation}\label{15}
\left(\dfrac{\partial S}{\partial m}\right)_\rho = -(p+\rho)/Tm^2.
\end{equation}

As $ T=T(\rho,m) $, the following identity becomes obvious:
\begin{equation}\label{16}
dT=\left(\dfrac{\partial T}{\partial m}\right)_\rho dm + \left(\dfrac{\partial T}{\partial \rho}\right)_m d\rho.
\end{equation}

Along with this we also have the integrability condition of the first law as
\begin{equation}\label{17}
T\left(\partial P /\partial \rho\right)_m = m \left(\partial T/ \partial m\right)_\rho + (p+\rho)\left(\partial T / \partial \rho\right)_m.
\end{equation}

The above two equations (\ref{16}) and (\ref{17}) can be solved for the unknowns $(\partial T/ \partial m)_\rho  $ and $ (\partial T / \partial \rho)_m $, and the condition for this is
\begin{equation}\label{18}
\dot{m}(p+\rho)-m\dot{\rho}=0.
\end{equation}

Now substituting this result back into the previous identities (\ref{16}) and (\ref{17}), we obtain the relation
\begin{equation}\label{19}
dT/T = (dm/m)(\partial P/\partial \rho)_m.
\end{equation}
If we now assume that the comoving particle number (proportional to $ma^3 $) of the fluid is conserved in the FRW universe, then we get the relation
\begin{equation}\label{20}
(\dot{m}/m)=-3 (\dot{a}/a),
\end{equation}
and substituting (\ref{20}) in the relation $$\dot{T}/T=(\dot{m}/m)\left(\dfrac{\partial p}{\partial \rho}\right)_m, $$ we obtain
\begin{equation}\label{21}
\dot{T}/T=-3 (\dot{a}/a)\left(\dfrac{\partial p}{\partial \rho}\right)_m.
\end{equation}
In order to determine $\left(\dfrac{\partial p}{\partial \rho}\right)_m  $, we use the equation of state of VMCG to arrive at the expression
\begin{equation}\label{22}
\left(\dfrac{\partial p}{\partial \rho}\right)_{m} =A +\frac{B_{0}\alpha a^{-n}}{\rho^{(\alpha + 1)}} + \frac{B_{0}n\rho^{-\alpha}}{a^{(n+1)}}\left(\dfrac{\partial a}{\partial \rho}\right)_{m}.
\end{equation}
Now using the conservation equation
\begin{equation}\label{23}
3\frac{\dot a}{a}(p+\rho)+ \dot \rho =0,
\end{equation}
and the equation of state for VMCG, we get
\begin{equation}\label{24}
\frac{B_{0}n\rho^{-\alpha}}{a^{(1+n)}}\left(\dfrac{\partial a}{\partial \rho}\right)_{m}=\frac{-nB_{0}a^{-n}}{3(1+A)\rho^{(1+\alpha)}-3B_{0}a^{-n}}.
\end{equation}
With the help of the equations (\ref{21}), (\ref{22}), (\ref{24}) and (\ref{09}) we finally obtain the following relation
%\begin{eqnarray}
%\nonumber
%\dfrac{dT}{T}=-3A\dfrac{da}{a}+ \\
%\nonumber
%\frac{3nB_0 \dfrac{da}{a}}{3(1+A)\rho_0^{1+\alpha}[\Omega_{x}+(1-\Omega_{x}a_0^{-n})(a_0/a)^{3N} a_0^n]-3B_0}
%\end{eqnarray}
%\begin{equation}
%- \frac{3B_0\alpha \dfrac{da}{a}}{\rho_0^{1+\alpha}[\Omega_{x}+(1-\Omega_{x}a_0^{-n})(a_0/a)^{3N} a_0^n]}
%\end{equation}
\begin{align}
&\left(\dfrac{dT}{T}\right) =-3A\left(\dfrac{da}{a}\right)+\nonumber \\
&\frac{3nB_0 \left(\dfrac{da}{a}\right)}{3(1+A)\rho_0^{1+\alpha}[\Omega_{x}+(1-\Omega_{x}a_0^{-n})(a_0/a)^{3N} a_0^n]-3B_0}\nonumber \\
&- \frac{3B_0\alpha \left(\dfrac{da}{a}\right)}{\rho_0^{1+\alpha}[\Omega_{x}+(1-\Omega_{x}a_0^{-n})(a_0/a)^{3N} a_0^n]}. \label{25}
\end{align}
From (\ref{25}), we now calculate the temperature as a function of scale factor $a(t)$. This yields
\begin{align}
\left.T(a)\right. &= T_{0}\left(\frac{1}{a}\right)^{3A}\left(\dfrac{1+(\frac{1-\Omega_{x}}{\Omega_{x}})}{a^{3N} +(\frac{1-\Omega_{x}}{\Omega_{x}}) }\right) ^{\frac{\alpha}{1+\alpha}} \nonumber \\
&\times\left( \dfrac{(\frac{n}{n-3N})+{(\frac{1-\Omega_{x}}{\Omega_{x}})}}{a^{3N}(\frac{n}{n-3N})+(\frac{1-\Omega_{x}}{\Omega_{x}})}\right).\label{26}
\end{align}

To derive the temperature $ T(z) $ as a function of the redshift $z$, we substitute $(z+1)=\frac{a_0}{a}$ in $ T(a)$, and finally arrive at the expression
%\begin{equation}
%\dfrac{dT}{T}=-3A\dfrac{da}{a}+ \frac{3nB_0 \dfrac{da}{a}}{3(1+A)\rho_0^{1+\alpha}[\Omega_{x}+(1-\Omega_{x}a_0^{-n})(a_0/a)^{3N} a_0^n]-3B_0} - \frac{3B_0\alpha \dfrac{da}{a}}{\rho_0^{1+\alpha}[\Omega_{x}+(1-\Omega_{x}a_0^{-n})(a_0/a)^{3N} a_0^n]}
%\end{equation}

% \begin{small}
% \begin{equation}
% T(z)=\frac{T_0(z+1)^{3N(1+\frac{\alpha}{1+\alpha}) +3A}(\frac{1}{\Omega_x})^{\frac{\alpha}{1+\alpha}}[\frac{1}{\Omega_x}(1-\frac{n}{3N})-\frac{1}{a_0^n}]a_0^{n(1+\frac{\alpha}{1+\alpha})}}{[1+(z+1)^{3N}(\frac{a_0^n}{\Omega_{x}}-1)]^\frac{\alpha}{1+\alpha}[(1-\frac{n}{3N})(\frac{a_0^n}{\Omega_{x}}-1)(z+1)^{3N}-\frac{n}{3N}]}
%\end{equation}
 %\end{small}
\begin{align} \label{T_z1}
\left. T(z)\right. &=\frac{T_0(z+1)^{3N(1+\frac{\alpha}{1+\alpha}) +3A}(\frac{1}{\Omega_x})^{\frac{\alpha}{1+\alpha}}}{[1+(z+1)^{3N}(\frac{a_0^n}{\Omega_{x}}-1)]^\frac{\alpha}{1+\alpha}} \nonumber \\
 &\times\frac{[\frac{1}{\Omega_x}(1-\frac{n}{3N})-\frac{1}{a_0^n}]a_0^{n(1+\frac{\alpha}{1+\alpha})}}
 {[(1-\frac{n}{3N})(\frac{a_0^n}{\Omega_{x}}-1)(z+1)^{3N}-\frac{n}{3N}]}.
\end{align}

\begin{figure}[ht]
 \centering
  \includegraphics[width=0.43\textwidth]{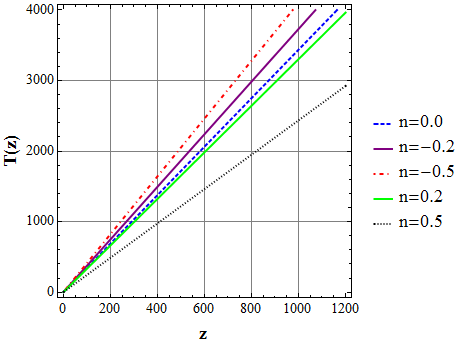}
 \caption{Variation of temperature  $ T(z) $ as a function of $z$ for different values of $ n $, where we have taken $ A=1/3 $, $ \Omega_x = 0.7 $, $ a_0=1 $, and $ \alpha=0.25 $ to see how the free parameter $ n $ affects the temperature $ T(z) $ in the VMCG model.}
\label{fig_1}
\end{figure}
This is the exact expression of temperature in terms of the redshift $ z $, for VMCG. In Fig.~\ref{fig_1}, we have plotted the temperature $ T(z) $ for different values of the free parameter $ n $. We can see that for large $ z $, the temperature decreases linearly with decreasing $ z $, but for small $ z $ it falls to zero in a gradual nonlinear fashion as $ z $ goes to negative values, indicating the possible future evolution of temperature of the universe. In this paper, wherever possible, we have extended the plots up to $z=-1$ in order to take into account the future evolution of the model.

\begin{figure}[ht]
\centering
\includegraphics[width=0.55\textwidth]{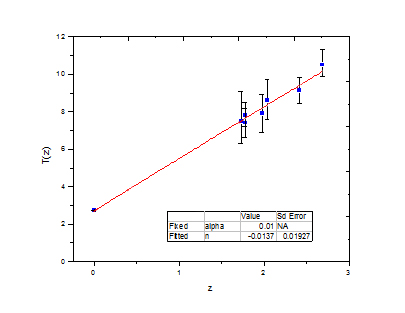}
\caption{A fit of the temperature $T(z)$ using some observational points available in literature (listed in TABLE I), where we have taken $ A=1/3 $, $ \Omega_x = 0.7 $, $ a_0=1 $, and $ \alpha=0.25 $ in the VMCG model. Here the value of $ \alpha $ was fixed to $ 0.01 $ and the fitted best value of $ n $ is $ -0.0137 $.}
\label{fig_1_b}
\end{figure}

The  cosmic microwave background radiation (CMBR) is a fundamental consequence of the hot Big-Bang. It is the radiation leftover after the decoupling from matter in the early evolutionary phases of the Universe. This radiation excites the rotational levels of some interstellar molecules, including carbon monoxide (CO), which can serve as a measuring device for the astronomers. Indirect measurement of $T(z)$ is one of the most powerful cosmological tests available.

Assuming that the CMB is the only source of excitation, Songaila et al. (1994) \cite{Songaila7.4} determined its temperature to be $T_{CMB} = 7.4 \pm 0.8 K$ from neutral carbon atoms at $z = 1.776$ in a cloud towards the quasar Q$1331+170$. Subsequent improvements have placed the estimate for the present CMB black-body temperature at the value of $T_{CMB} = 2.725\pm 0.002 K $ (Mather et al. 1999 \cite{mathercobe}), which was measured locally (at redshift z = 0).

Lima et al. \cite{lima} argued that the CMB temperature at high $z$ may be smaller than the predicted standard values, which opened the scope of alternative models for the big bang. Using new elements in the form of decaying vacuum energy density and gravitational `adiabatic' photon creation along with the late inflationary models driven by a scalar field, they deduced a new temperature law and compared its predictions with the standard cosmological results.

Srianand et al. (2008) \cite{sri9.15}, also assumed the CMB to be the only source of excitation and imposed stringent upper-limits on $T_{CMB}$ for a large sample of C {\small I} fine structure absorption lines detected in high signal to noise, high resolution spectra. They detected carbon monoxide in a damped Lyman$-\alpha$ system at $z_{abs} = 2.41837$ in the SDSS database towards SDSS J$143912.04+111740.5$, and from the CO excitation temperatures they determined $T_{CMBR} = 9.15 \pm 0.72 K$.

In their paper, J. Ge et al. \cite{BECHTOLD7.9} presented the detection of absorption lines from the ground state and excited states of C {\small I} in the $z=1.9731$ damped Lyman$\alpha$ system of the QSO $0013-004$ and estimated other contributions to the excitation of the C {\small I} fine-structure levels. They used the population ratio of the excited state to the ground state and estimated the CMBR temperature of $T =7.9 \pm 1.0 K$ at $0.61$ mm and $z=1.9731$, which matched with the predictions of standard cosmology at that time.

Noterdaeme et al. (2010) \cite{Noterdaeme10.5}, in  their paper, presented the analysis of a sub damped Lyman-$\alpha$ system with neutral hydrogen column density at $z_{abs}=2.69$ toward SDSS J$123714.60 + 064759.5$. The excitation of CO was found to be dominated by radiative interaction with the CMBR and they derived $T_{ex}(CO) = 10.5 K$ corresponding to the expected value of $T_{CMBR}(z=2.69)=10.05 K $.

Using three new and two previously reported CO absorption line systems detected in quasar spectra during a
systematic survey carried out using VLT$/$UVES, P. Noterdaeme et al. \cite{ Noterdaeme} constrained the
evolution of $T_{CMB}$ to $z\sim3$. Combining their measurements with previous constraints, they obtained $T_{CMB}(z)=(2.725 \pm 0.002)\times(1 + z)^{1-\beta} K$ with $\beta = -0.007\pm 0.027$.

All these have motivated us to derive exact expression for the temperature of the FRW universe dominated by VMCG matter as a function of redshift $z$, in order to check for the viability of this cosmological model by using this observational constraint.

\begin{table}[htbp]
  \centering
  \caption{ T(z) table for different values of redshift as obtained from different references mentioned in the column}
    \begin{tabular}{|rrr|}
    %\toprule
    \hline
    z \,&\, T(z) \,&\, Reference \\
     \hline
      \hline
    1.776 \,&\, $7.4_{-0.8}^{+0.8}$ \,&\, \cite{Songaila7.4}  \\
    1.7293 \,&\, $7.5_{-1.2}^{+1.6}$ \,&\, \cite{Noterdaeme}\\
    1.7738 \,&\, $7.8_{-0.6}^{+0.7}$ \,&\, \cite{Noterdaeme}\\
    2.6896 \,&\, $10.5_{-0.6}^{+0.8}$ \,&\, \cite{Noterdaeme10.5,Noterdaeme}\\
    2.4184 \,&\, $9.15_{-0.7}^{+0.7}$ \,&\, \cite{sri9.15,Noterdaeme}\\
    2.0377 \,&\, $8.6_{-1.0}^{+1.1}$ \,&\, \cite{Noterdaeme}\\
    1.9731 \,&\, $7.9_{-1.0}^{+1.0}$ \,&\, \cite{BECHTOLD7.9}\\
    0     \,&\, $2.725_{-0.002}^{+0.002}$ \,&\, \cite{mathercobe}\\
    %\bottomrule
    \hline
    \end{tabular}%
  \label{tab:addlabel}%
\end{table}%

We have used some of the observational temperature data points for different redshifts in the Fig.\ref{fig_1_b}  from the Table \ref{tab:addlabel} and used our theoretical curve to show the overall agreement with the cosmological observations.

%\begin{eqnarray}
 % T(z) &=& \dfrac{T_{0}(z+1)^{7-\dfrac{6n}{5}}(\dfrac{1}{\Omega_{x}})^{1/5} a_{0}^{6n/5}}{[1+(z+1)^{5-n}(\dfrac{a^{n}_{0}}{\Omega_{x}}-1)]^{1/5}}  \nonumber \\
  % & & \times \dfrac{ (5-2n)\Omega_{x}^{-1} - (5-n){a_{0}^{-n}} }{(5-2n)((a_{0}^{n})\Omega_{x)^{-1} - 1)(z+1)^{5-n} - n}
%\end{eqnarray}
If we substitute the parameter values $\alpha=0.25$, $A=1/3$ and $ \Omega_{x}=0.7$ in (\ref{T_z1}), then we get the expression of temperature as a function of the free parameter $n$:
%\begin{equation}
%T(z)=T_0(z+1)^{3(R-1)}[\Omega_x + (1-\Omega_x)(z+1)^{3R}]^{-\alpha/1+\alpha}
%\end{equation}
\begin{align}
 T(z) &=2.9(z+1)\left[(z+1)^{n-5}+\frac{3}{7}\right]^{-1/5} \nonumber \\
 & \times \left(\dfrac{(13n-15)}{7n(z+1)^{n-5}+6n-15}\right).\label{28}
\end{align}
We should be able to get the corresponding expression for MCG if we put $n=0$ in equation (\ref{28}). After substituting $n=0$ in (\ref{T_z1}), we obtain
\begin{align}
T(z) =& T_0(z+1)^{3(R-1)}\nonumber \\
 & \times [\Omega_x + (1-\Omega_x)(z+1)^{3R}]^{-\alpha/1+\alpha}. \label{29}
\end{align}
This result matches exactly with the corresponding expression for the MCG as calculated by Bedran et al \cite{MCG3}. We now have a working formula for the temperature. We can use the boundary conditions as $ T_0=2.7K $ and substitute the values of other variables like $ A=1/3 $ (for the radiation phase) and $ \Omega_x = 0.7 $ (most commonly used and accepted dark energy parameter) in equation (\ref{T_z1}) to get the expression of temperature in terms of $ \alpha $ and $n$. If we substitute $ n=0 $ (the MCG case) and $ \alpha=\frac{1}{4} $ in the resulting expression, then we obtain
\begin{equation}
T(z)=2.90K (z+1)[(z+1)^{-5}+3/7]^{-\frac{1}{5}},\label{30}
\end{equation}
which is consistent with the expression of temperature in the paper \cite{MCG3}. Unfortunately for VMCG we don't know the constraints on $ \alpha $ and $ n $. From the consideration of thermodynamic stability in the case of VMCG, one of the authors \cite{VMCG2} have shown that the condition for stability is $ n\leq 0 $.

\begin{figure*}[ht]
 \centering
  \includegraphics[width=0.9\textwidth]{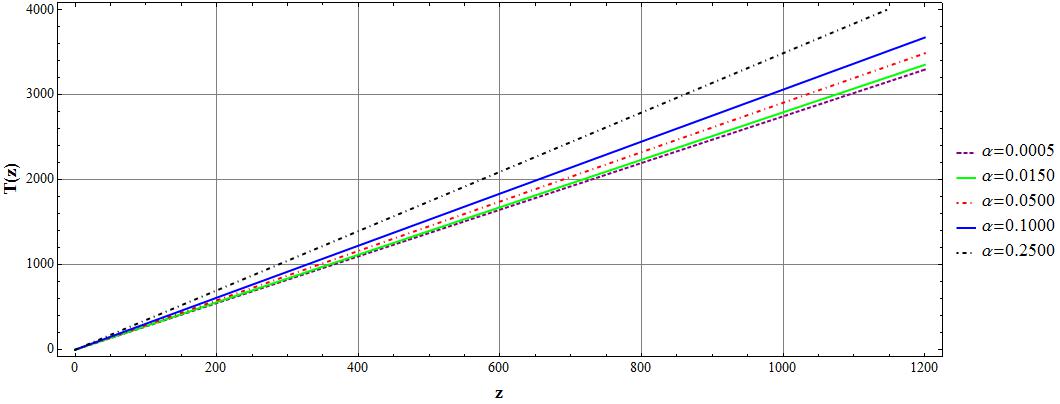}
 \caption{Variation of temperature  $ T(z) $ as a function of $z$ for different values of $ \alpha $, where we have taken $ A=1/3 $, $ \Omega_x = 0.7 $, and $ n=-0.01 $ to see how the free parameter $ \alpha $ affects the temperature $ T(z) $ in the VMCG model.}
\label{fig_1a}
\end{figure*}
In Fig.~\ref{fig_1a} we have shown the temperature evolution of the universe as a function of redshift for different values of $ \alpha $ where we have chosen a fixed value of the parameter $ n=-0.5 $. We can clearly see that as the value of $ \alpha $ increases, the temperature increases for a particular value of $n $. For the MCG model, from the above equation (\ref{30}), using the decoupling redshift as $ z\approx1100 $, the temperature of decoupling ($T_{d}$) is found to be $ T_{d}\approx 3800K $. For VMCG the decoupling temperature is complicated, and depends on the parameters $ n$ and $\alpha $. Putting the values $ A=1/3 $, $z=1100$ and $\Omega_{x}=0.7$ in equation (\ref{T_z1}), we arrive at the following relation
\begin{align} \label{T_d}
\left. T_{d}(n,\alpha)\right. &=\frac{(2.7)(1101)^{3N(1+\frac{\alpha}{1+\alpha}) +1}(1.43)^{\frac{\alpha}{1+\alpha}}}{[1+(1101)^{3N}(0.43)]^\frac{\alpha}{1+\alpha}} \nonumber \\
 &\times\frac{[(1.43)(1-\frac{n}{3N})-1]}
 {\left[(1-\frac{n}{3N})(0.43)(1101)^{3N}-\frac{n}{3N}\right]}.
\end{align}
where $3N=4(1+\alpha)-n$.

\begin{figure}[ht]
 \centering
  \includegraphics[width=0.40\textwidth]{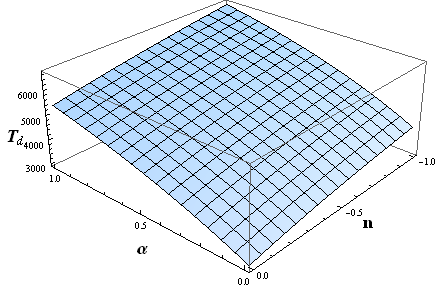}
 \caption{Variation of decoupling temperature $T_{d}$, as a function of $n$ and $\alpha$, where we have taken $ A=1/3 $, $ \Omega_x = 0.7 $, $ a_0=1 $ and $ z=1100 $ to see how the two free parameters affect the decoupling temperature.}
\label{fig_2}
 \end{figure}

In Fig.~\ref{fig_2} we have shown the dependence of decoupling temperature on the parameters $ n $ and $ \alpha $ for the VMCG model using the expression for decoupling temperature in equation (\ref{T_d}). We have calculated the values of decoupling temperatures for the VMCG model using $ \alpha=1/4 $, and $ n=-0.1 , -1.0 , -2.0 $, and the corresponding values are $ T_{d}\approx $ 3941 K, 5033 K, and 5764 K respectively. Thus we can see that the decoupling temperature increases with higher negative values of $ n $ for a fixed value of $ \alpha $ in the VMCG model.

The expression for energy density of VMCG in FRW universe can be used to determine the Hubble parameter $ H=\frac{\dot a}{a} $ as a function of redshift $ z $. We know that for the flat FRW universe we have
\begin{equation}
3({\dot a}/{a})^2 = \rho,
\end{equation}
which gives $ H^2=\rho/3 $, and so we have
\begin{equation}
H^2=H_{0}^2 (\rho/\rho_{0}).
\end{equation}
Now using the expression for $ \rho(a) $ and changing our variable to the redshift $ z $, we get
\begin{align}
H(z)= & H_{0}(z+1)^{n/2(1+\alpha)}\nonumber \\
 & \times [\Omega_{x} + (a_{0}^n - \Omega_{x})(1+z)^{3N}]^{1/2(1+\alpha)}.
\end{align}

\begin{figure}[ht]
\centering
  \includegraphics[width=0.4\textwidth]{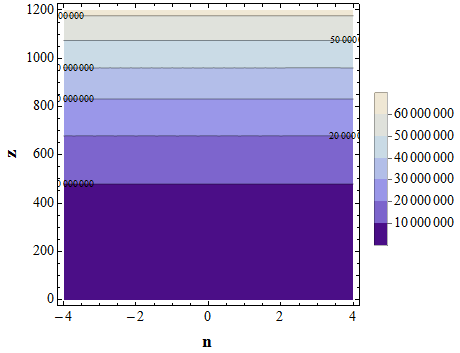}
\caption{Contour plot of Hubble parameter $H(z)$ as a function of $ z $ and $n$, where we have taken $ H_0=70 $, $ A=1/3 $, $ \Omega_x = 0.7 $, $ a_0=1 $, and $ \alpha=0.25 $ to see how the free parameter $ n $ affects $ H(z) $ in the VMCG model.}
\label{fig_3}
\end{figure}

In the contour plot of Fig.~\ref{fig_3} we have shown how the Hubble parameter varies with $ z $ for different values of $ n $. We can see that for high values of $ z $ there is no significant shift in the magnitude of the Hubble parameter for different values of $ n $.

\begin{figure}[ht]
 \centering
  \includegraphics[width=0.45\textwidth]{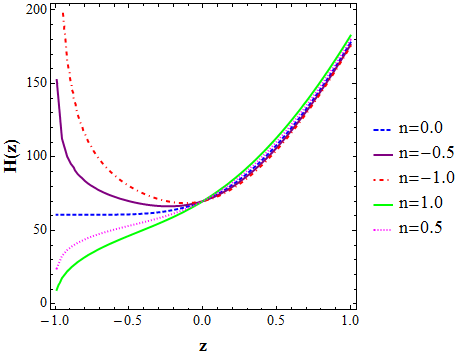}
 \caption{Plot of the Hubble parameter $H(z)$ as a function of $ z $ and $n$, where we have taken $ H_0=70 $, $ A=1/3 $, $ \Omega_x = 0.7 $, $ a_0=1 $, and $ \alpha=0.25 $ to see how the free parameter $ n $ affects $ H(z) $ in the VMCG model.}
 \label{fig_4}
 \end{figure}

We can also see in Fig.~\ref{fig_4} that for positive $ n $ the value of $ H $ decreases and approaches zero, whereas for negative $ n $ (i.e. phantom dominated universe) it increases rapidly as $ z $ approaches $ -1 $, indicating the Big rip that will occur in future. As the square of the Hubble parameter is proportional to the energy density $ \rho(z) $, it is clear that for negative $ n $ (thermodynamically stable condition) the energy density increases rapidly to infinity, as it should be in a phantom dominated universe as $ z $ approaches $ -1 $. In Fig.~\ref{fig_4} and in the subsequent figures, we have included the range $ -1<z<0 $ (which indicates blueshift with respect to the present epoch) to show how the different parameters of this VMCG model, like $H(z),\, z,\, W(z),\, q(z)\, \textrm{and}\, v^2(z)$, will vary in the future, and how they will differ from each other depending on the values of the other parameters in this model.

% same thing have to be done fixing n and varying alpha . also using the exact expression of T we will show in different limits how the teparature relation varies with other parameters .

\begin{figure}[ht]
 \centering
  \includegraphics[width=0.45\textwidth]{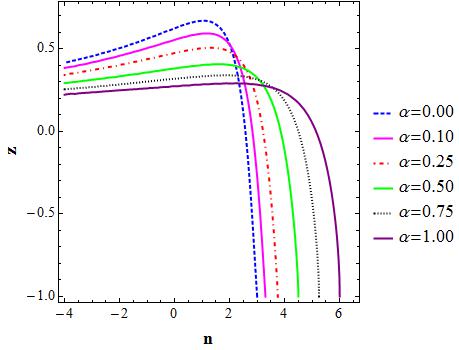}
 \caption{Variation of the redshift $z$ (for $ P=0 $) as a function of $n$, where we have taken $ A=1/3 $, $ \Omega_x = 0.7 $, $ a_0=1 $, $ \alpha=0.25 $ to see how the free parameter $ n $ affects the redshift of the dust phase in the VMCG model.}
 \label{fig_5}
\end{figure}

We can also find the redshift $ z $ when the pressure passes through the zero value ($ P=0 $). We substitute $ P=0 $ in the equation of state and use the expression for $ \rho(a) $ and substitute $ B_0 $ in terms of $ \Omega_x $ to arrive at the following expression:
\begin{equation}
[\Omega_x + (1-\Omega_x)(z+1)^{4(1+\alpha)-n}]= \left[\dfrac{4(1+\alpha)-n}{(1+\alpha)}\right]\Omega_{x}.
\end{equation}
With $ A=1/3 $, $ \Omega_x = 0.7$ and setting $ n=0 $ (for the MCG) and $ \alpha = 1/4 $, we obtain the redshift as $ z=0.48 $. For nonzero values of $ n $, we get (for $ P=0 $) the dependence of $ z $ on $ n $ for different values of $ \alpha $ as shown in the Fig.~\ref{fig_5}.

It is evident that as we move towards more negative values of $ n $, the value of redshift (for $ P=0 $) decreases very slowly. As we vary $ \alpha $, we can see that as it increases for a fixed negative value of $ n $, the value of $ z $ decreases. Therefore, for a phantom dominated universe in the VMCG model, the redshift for dust phase must be $ z<0.48 $ for the chosen values of the parameter $ \alpha = 0.25 $, whereas for positive $ n $ there is no such bound.

\begin{figure}[ht]
 \centering
  \includegraphics[width=0.45\textwidth]{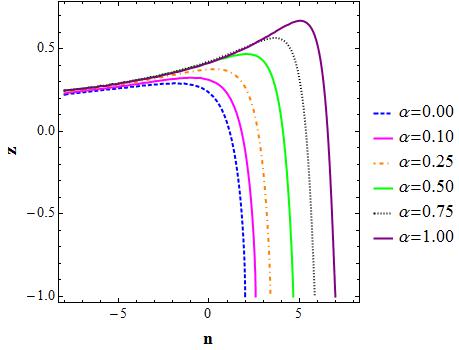}
 \caption{Variation of redshift (for $ \ddot{a}=0 $) as a function of $n$, where we have taken $ A=1/3 $, $ \Omega_x = 0.7 $, $ a_0=1 $ and $ \alpha=0.25 $ to see how the free parameter $ n $ affects the redshift at the time of flip in acceleration in the VMCG model.}
 \label{fig_6}
 \end{figure}

We can also calculate the redshift of transition of the expansion of the universe from deceleration to acceleration. Using the equation of state and setting the condition $ \ddot{a}=0 $, i.e. $ 3P+\rho=0 $, we get the following relation
\begin{equation}
[\Omega_x + (1-\Omega_x)(z+1)^{4(1+\alpha)-n}]= \left[\dfrac{4(1+\alpha)-n}{2(1+\alpha)}\right]\Omega_{x},
\end{equation}
where we have assumed that $ A=1/3 $. Now if we substitute $ \Omega_x = 0.7$, $ \alpha=0.25 $ and $ n=0 $ (representing the MCG), we get $ z=0.18 $ (for MCG). The variation of $ z $ (for transition from deceleration to acceleration phase) as a function of $ n $ for different values of $ \alpha $ can also be seen in Fig.~\ref{fig_6}.

As we vary $ \alpha $, we can see that as it increases for a fixed negative value of $ n $, the value of $ z $ increases. From the plot it is clear that if $ n>0 $ (i.e. the Big rip is avoided), the value of the redshift for the flip in acceleration in the VMCG model must be $ z<0.18 $ for the chosen values of the parameter $ \alpha=0.25 $, but for negative value of $ n $, such a conclusion cannot be drawn.

\section{Discussions}

\begin{figure}[ht]
 \centering
  \includegraphics[width=0.45\textwidth]{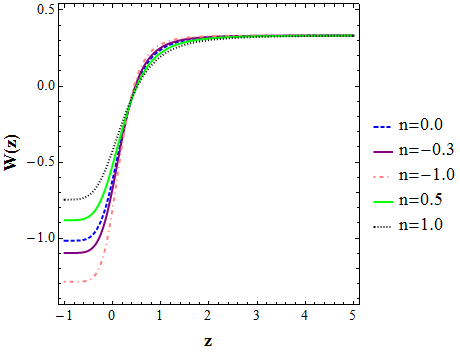}
 \caption{Variation of EOS parameter $\textit{W(z)}$ for different values of $n$, where we have taken $ A=1/3 $, $ \Omega_x = 0.7 $, $ a_0=1 $ and $ \alpha=0.25 $ to see how the free parameter $ n $ affects $\textit{W(z)} $ in the VMCG model.}
 \label{fig_7}
 \end{figure}

In this section we discuss the variation of some useful cosmological parameters of the VMCG in terms of the redshift $ z $, which is a parameter which can be easily measured through observations. First we consider the equation of state (EOS) for the VMCG and use the expression of $ \rho(a) $ to derive the equation of state parameter $ \textit{W(z)}=P/\rho $ in the form
\begin{equation}\label{W_z}
 \textit{W(z)}=A - \frac{N}{(1+\alpha)}\frac{1}{[1+(\frac{1}{\Omega_{x}}-1)(z+1)^{3N}]}.
\end{equation}
Analysing (\ref{W_z}) we find that for high $ z $, the EOS parameter approaches $\textit{W(z)}\simeq A$, and as it should correspond to the `Radiation-dominated phase' of the universe, we can safely say that $ A $ must have the value $ 1/3 $. For small $ z $, the EOS parameter approaches the value $ \textit{W(z)}\simeq -1+\frac{n}{3(1+\alpha)} $. As this expression explicitly depends on $ n $, it means that if $ n $ is negative, then $\textit{W(z)}<-1 $, which corresponds to the phantom-dominated universe and Big rip is unavoidable, whereas for $ n\geq 0 $, the EOS parameter becomes $ \textit{W(z)}\geq -1 $, so that Big rip is avoided. From the plot in Fig.~\ref{fig_7} we find that for different values of $ n $, as $ z $ increases, the value of the EOS parameter approaches $1/3$, and further the plot also shows the position where the pressure becomes zero and then negative, approaching different negative values for different values of $ n $. The value of the redshift for the `dust phase' ($ P=0 $) is very close to the value of $ z\simeq 0.48 $ depending on the value of $ n $, which agrees with our analysis in the previous section.

\begin{figure}[ht]
 \centering
  \includegraphics[width=0.45\textwidth]{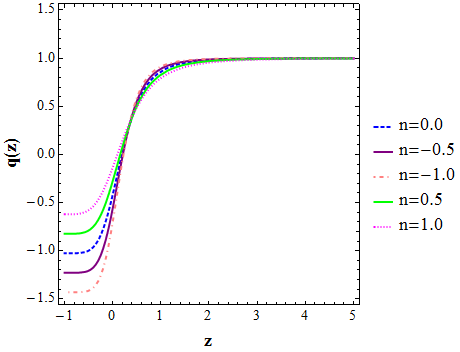}
 \caption{Variation of deceleration parameter ($q(z) $) for different values of $n$, where we have taken $ A=1/3 $, $ \Omega_x = 0.7 $, $ a_0=1 $ and $ \alpha=0.25 $ to see how the free parameter $ n $ affects $q(z) $ in the VMCG model.}
 \label{fig_8}
 \end{figure}

It is known that the deceleration parameter $ q(z) $ is related to $\textit{W(z)}$ by the relation \cite{VMCG2}
\begin{equation*}
 q(z)=1/2 + 3\textit{W(z)}/2.
\end{equation*}
Thus the expression for the deceleration parameter $ q(z)  $ as a function of redshift $ z $ is
\begin{align}
q(z)= & 1/2 + (3/2) \nonumber \\
 & \times \left[A - \frac{N}{(1+\alpha)}\frac{1}{[1+(\frac{1}{\Omega_{x}}-1)(z+1)^{3N}]}\right].
\end{align}
Fig.~\ref{fig_8} clearly indicates the variation of $q(z)$ with $ z $ for different values of $ n $. For large $ z $, the expression becomes  $ q(z)\simeq 1/2 + 3A/2 $, which is constant, and for small $ z $ the deceleration parameter takes the form $ q(z)\simeq -1 + \frac{n}{2(1+\alpha)} $, which again explicitly depends on $ n $. This means that $q(z)$ was constant in the radiation phase, then gradually decelerated while passing through the dust phase and then entered the current accelerating dark energy dominated phase. Depending on the value of $ n $, $q(z)$ approaches different values for small $ z $. For positive values of $ n $, $ q(z)>-1 $ and for $ n \leq 0 $, $ q(z) \leq -1 $. From the figure one can easily see that $q(z)$ crosses zero (i.e. the moment of flip from deceleration to acceleration) near $ z\simeq 0.18 $, depending on the value of $ n $, which conforms to our calculations in the previous section.

Similarly we can calculate the velocity of sound $ v^2_{s}=(\partial P/\partial \rho)_{s} $. Using the equation of state for VMCG we obtain the relation
\begin{equation}\label{v_s}
v^2_{s} = A +\frac{B\alpha}{\rho^{(1+\alpha)}}-\frac{\rho^{-\alpha}B_{0}(-n)a^{-(n+1)}}{\partial \rho/\partial a}.
\end{equation}
If we calculate $(\partial \rho/\partial a) $ from (\ref{09}) and substitute it in (\ref{v_s}), then we get the expression for the velocity of sound as a function of $ z $:
\begin{align}
\left.v^2_{s}(z)\right. &= A + \frac{(N\alpha/(1+\alpha))}{[1+(\frac{1-\Omega_{x}}{\Omega_{x}})(z+1)^{3N}]} \nonumber \\
&-\frac{Nn}{n +(\frac{1-\Omega_x}{\Omega_x})(1+z)^{3N}(n+3N)}
\end{align}

\begin{figure}[ht]
 \centering
  \includegraphics[width=0.45\textwidth]{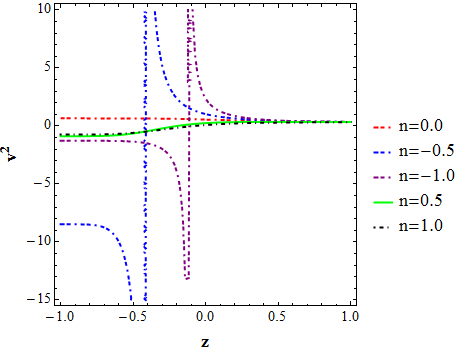}
 \caption{Variation of velocity of sound $ v^2_{s}(z) $ for different values of $n$, where we have taken $ A=1/3 $, $ \Omega_x = 0.7 $, $ a_0=1 $ and $ \alpha=0.25 $ to see how the free parameter $ n $ affects $v^2_{s}(z) $ in the VMCG model.}
 \label{fig_9}
\end{figure}

From Fig.~\ref{fig_9} we find that the velocity has a magnitude lying below unity and the nature is consistent at large $ z $, because for large value of redshift i.e., in the early phase of the universe, the velocity was  $v^2_{s}(z)\simeq A =1/3  $, and then as the redshift $ z $ became smaller, the velocity of sound increased rapidly for $ n<0 $ (phantom dominated universe). After that the velocity becomes imaginary. Whereas for $ n>0 $, as the redshift decreases, it slowly decreases and becomes negative. For small values of $ z $, the velocity of sound is given by $ v^2_{s}(z)\simeq -1+ \frac{n}{3(1+\alpha)} $, which for negative $ n $ is always negative (signifying imaginary speed), and for the speed to be real, we must have $ n>3(1+\alpha) $, which is thermodynamically unstable for positive $ \alpha $. Therefore this scenario for $ n<0 $ indicates a perturbative cosmology and favours structure formation in the universe \cite{sound}. Whereas for $ n=0 $, the velocity approaches a constant positive value for small values of $ z $.

\section{Conclusions}

In this paper, we have determined the energy density of VMCG matter in a FRW universe using a thermodynamic approach, and derived the exact expression for the temperature $ T(z) $ and the Hubble parameter $ H(z) $ of the corresponding universe as a function of redshift $ z $. We have derived the redshift for the `dust phase' ($ P=0 $) and for  the epoch of transition from deceleration to acceleration ($ \ddot{a}=0 $) of the FRW universe. We have also determined the dependence of the redshift during the different phases of expansion of the universe on the free parameter $ n $. Subsequently we have shown the dependence of the equation of state parameter $ \textit{W(z)} $, deceleration parameter $ q(z) $ and the velocity of sound $ v^2(z) $ on the free parameter $ n $ as a function of redshift $z$. We find that the VMCG model perfectly represents the three different phases of the universe namely the `Radiation phase' ($ P=\rho/3 $), the `dust phase' ($ P=0 $), and later the negative pressure epoch dominated by the so called `Dark energy'. We also find that the VMCG model with $ n<0 $ (for thermodynamic stability) and other accepted values of the parameters, explains the value of decoupling temperature very well.

Therefore from the above analysis we can conclude that a FRW universe filled with thermodynamically stable variable modified Chaplygin gas not only represents the three phases of evolution of the universe very well along with the change of expansion rate from deceleration to acceleration, but also it shows a consistent temperature evolution of universe.

In this context we like to mention that two of the authors have also examined the validity of the generalized second law of thermodynamics (GSLT) on the cosmological apparent horizon (AH) and the event horizon (EH) of FRW universe dominated by various types of Chaplygin gas fluids, one of them being the VMCG \cite{ChakGuha}. The GSLT is always valid on the apparent horizon of the VMCG dominated FRW universe. But for $ n<0 $ (i.e. $R_{EH}<R_{AH}  $), the VMCG dominated FRW universe violates the GSLT on the event horizon in the early phase of the universe but it holds in our current epoch and will also hold in the future, where we know that in this range of $ n<0 $, the VMCG model itself is thermodynamically stable \cite{VMCG2}.

\section*{Acknowledgments}
We wish to thank CSIR, Government of India, for financial support. A portion of this work was done in IUCAA, India. Both SC and SG are grateful to IUCAA for the warm hospitality and the facilities of work available there.

\end{document}